# A New Approach to the Glass Transition of Percolated Polymers from the Perspective of Thermal Volume Expansion

Jan-Kristian Krüger*, Bernd Wetzel, Andreas Klingler*

Leibniz-Institut für Verbundwerkstoffe (IVW), 67655 Kaiserslautern, Germany

**Summary Paragraph.** *In 1995, the Nobel Price winner P.W. Anderson made the following remarkable statement: "The deepest and most interesting unsolved problem in solid state theory is probably the theory of the nature of glass and glass transition."[1]. Although there have been new theoretical developments in the meantime, in our opinion, this situation has only improved marginally to date. One of the main reasons is the insufficient consideration of experimental boundary conditions. A central experimental problem arises from the fact that the time constants required to achieve thermodynamic equilibrium increase sharply in the vicinity of and above a hypothetical static glass transition. If these equilibrium conditions are violated, additional internal thermodynamic variables come into play that normally alter the static and dynamic susceptibilities significantly and thus lead to misinterpretations of the experimental data. This raises, for example, the question of whether there are static property changes during the canonical glass transition and how these correlate with dynamic precursors. In this article, we attempt to find answers to these questions using a new experimental method called temperature-modulated optical refractometry in combination with the temperature-jump technique. The total time required to investigate the glass transition behaviour of our model epoxy was approximately two years.*

At least since the legendary publications by Kovacs and colleagues[3–5] on thermally induced, kink-like volume changes during the canonical glass transition of polymers, such volume changes have been regarded as an important feature of the canonical glass transition[6,7]. The kink temperature $T_g$ is usually assumed to be the operative transition temperature. According to Kovacs, this $T_g$ depends on the cooling rate[4]. The specific volume measurements of Kovacs et al. in the vicinity of the canonical glass transition of polyvinyl acetate (PVAc)[4] were performed temperature-rate dependent on cooling. As a consequence, the resulting volume data $V(T)$ and the thermal volume expansion coefficient $\beta(T) = 1/V \cdot dV/dT$ were neither measured under thermal equilibrium conditions - i.e. they were dynamic in nature – nor were they measured as a function of frequency $\nu$, i.e. the non-equilibrium state was created through the temperature rate but not a modulation frequency.

In order to obtain a clearer view on the freezing dynamics given by the so-called $\alpha$-process and its convergence versus a hypothetical static glass transition[8–10], thermal volume changes due to temperature changes together with frequency changes of the complex dynamic susceptibility $\beta^*(T,\nu)$ are desirable with the goal to realize $\beta_{stat}(T) = \lim_{\nu \to 0} \beta^*(T,\nu)$. According to Baur[11], such measurements of the thermal volume expansion $\beta^*(T,\nu)$ should be performed in linear response approximation of irreversible thermodynamics. However, the implementation of this condition is not trivial. First, the glass precursor in question should be in thermodynamic equilibrium, even near the static glass transition ($T > T_{gs}$), without external temperature disturbance. Second, the measuring system should be stable enough in the long term that even measurements in the µHz range can still provide statistically relevant data. And third, of course, it must be ensured that all measurements meet the conditions of linear response in terms of equilibrium thermodynamics[12].

In order to better understand the nature of the canonical glass transition, particularly in polymers, it is desirable and necessary to gain a more accurate understanding of the coupling between static and dynamic thermal volume expansion properties. An important prerequisite to realize all these conditions is the simultaneous determination of the static and dynamic properties of thermal volume expansion. A particularly suitable measurement method for this is “temperature-modulated optical refractometry” (TMOR, Anton Paar OptoTec GmbH, Seelze, Germany)[2,13–15]. In combination with the temperature jump method[16–18] TMOR meets all these conditions under cooling from a viscoelastic to a solid, glassy state at modulation frequencies down to a few µHz. Because even under the strictest thermodynamic measurement conditions, kinetic influences on the measured susceptibilities cannot be completely avoided, the term s*tatic* glass transition temperature $T_{gs}$ will in the following be replaced by the term *quasi-static* transition temperature $T_{gqs}$. In the Method section, we provide a detailed discussion and evaluation of the implementation of the three conditions in the current case[13].

However, the biggest challenge in this context is to determine the intrinsic complex thermal volume expansion coefficient $\beta^*(T,\nu)$ under thermodynamic quasi-equilibrium conditions, i.e., to eliminate all influences imposed by a temperature rate on approaching $T_{gqs}$. The following experiments were performed using a percolated model DGEBA/DETA polymer. See the Methods section for preparation details.

The thermally induced volume changes in the vicinity of the canonical glass transition of our model epoxy are determined in this publication using the TMOR method. For the sake of clarity, the basic equations are given in the following. An extensive introduction to TMOR can be found elsewhere[2,13–15].

The TMOR method is an optical measurement method that uses the Lorenz-Lorentz equation[19,20] to provide access to the static thermal volume expansion coefficient $\beta$, cf. Ref.[2].

$$\beta(T) = \frac{-6 \cdot n(T)}{(n^2(T)-1)(n^2(T)+2)} \cdot \frac{\partial n(T)}{\partial T} \tag{1}$$

where $n$ is the refractive index and corresponds to the usual $n_D$ ($\lambda = 589\ nm$). Provided, that the temperature jump method brings the sample material into thermodynamic equilibrium after each temperature change, Eq. 1 allows the quasi-static thermal volume expansion $\beta$ to be determined, if the temperature jumps are sufficiently small. If the temporal temperature disturbance $T(t)$ includes an additional modulation signal

$$T(t) = T_0 + A_T \cdot \sin(2 \cdot \pi \cdot \nu \cdot t), \qquad (2)$$

the frequency-dependent (complex) thermal volume expansion coefficient $\beta^* = \beta' - i\beta''$ can be determined using TMOR[13].

$$\beta'(\nu) \equiv Re(\beta^*_{2\pi\nu}) = |\beta^*_{2\pi\nu}| \cos\Phi \qquad (3a)$$

and

$$\beta''(\nu) \equiv Im(\beta^*_{2\pi\nu}) = |\beta^*_{2\pi\nu}| \sin\Phi \qquad (3b)$$

where $\Phi$ is the phase angle between the exciting temperature perturbation and refractive index response.

At this point, it is important to note that using TMOR additional static $\beta$-data are generated using temperature modulation, provided, of course, that the phase angle $\Phi$ in Eqs. 3a and 3b essentially disappears. This discussion will play an important role later when we discuss $(\beta', \beta'')$, cf. *Methods section*. In this context, it is important to understand that the TMOR method can be used to determine thermal volume expansion coefficients under quasi-isothermal conditions because the necessary temperature disturbance is provided by a sufficiently small temperature modulation. This experimental possibility is a prerequisite for the important application of the temperature jump method.

Correct measurements of susceptibilities require the *linear response* condition to be guaranteed. Reliable TMOR investigations therefore require a priori control of the amplitude parameter $A_T$ in Eq. 2, as this is the only way to ensure the required linear response during TMOR investigations[13]. Extensive, preliminary investigations[13], amongst others in the vicinity of the canonical glass transition, have shown that the linear response condition is not violated using temperature modulation amplitudes up to $A_T = 0.3\ K$. In this study, to ensure linear response conditions, $A_T$ was selected to be as small as 0.1 K.

Avoiding cooling rate effects during temperature jumps, that eventually provide thermal equilibrium conditions, are a real challenge[13]. The temperature jump method was introduced about 20 years ago in connection with low-frequency time-domain Brillouin (TDB) spectroscopy[21]. At that time, the aim was not to generate thermodynamic equilibrium states as close as possible to the canonical glass transition. The aim at that time was to use TDB spectroscopy to achieve extremely low-frequency relaxation times in the range of the so-called $\alpha$-process[22], i.e., on the Vogel-Fulcher-Tamann (VFT)[23–25] diagram, and that procedure was successful. Today, using time- and frequency-dependent TMOR spectroscopy, it is possible to roughly estimate the time required to establish thermodynamic equilibrium near the canonical glass transition based on the VFT behaviour of our model epoxy. The extreme curvature of the VFT curve near the canonical glass transition, i.e., the enormous and disproportionate increase in the $\alpha$-relaxation times, gives an idea of the time required for the approximate establishment of thermodynamic equilibrium near $T_{gqs}$.

Our experimental approach for the temperature jump method is presented and discussed in more detail in Fig. 5 to Fig. 9, in the Methods sections. The cooling experiment begins well above the estimated glass transition temperature in the viscoelastic state in internal equilibrium, when the derivatives of the free energy density with respect to all relevant internal variables disappear relatively quickly[11,26]. Starting from this initial temperature regime, the temperature is lowered in sufficiently small steps to enable rapid equilibration of the thermodynamic system.

Fig. 1 shows TMOR investigations of the short time average refractive index $\langle n \rangle$, realized already in the TMOR instrument, as a function of temperature $T$. Using the Lorenz-Lorentz Eq. 5, we estimate a normalized specific volume $\langle v \rangle$, which gives a rough overview of the glass transition behaviour of our model epoxy as a function of temperature. The so-called *operational* glass transition temperature $T_g$ was calculated as usual by the intersection of the high-temperature ($g_1$) and low-temperature fit lines ($g_2$) of $\bar{n}(T)$ at $T_g = 31.3\ °C$. In this case, the temperature $T_g$ has no well defined physical meaning. The investigations were performed with a modulation time of 60 s ($\nu = 17\ mHz$). The dynamic TMOR measurements in Fig. 1 where all performed on cooling from the visco-elastic to the glassy state with the a rather small modulation amplitude $A_T = 0.1\ K$. The probed temperature range starts well above the glass transition in the visco-elastic range and ends in the glassy state. The normalized specific volume $\langle v \rangle$ was calculated using the ambiguous assumption $r = 1$ (cf. again the Methods section). The temperature dependence of $\langle n \rangle = \bar{n}$ in Fig. 1 gives a rough idea about the temperature dependence of the prefactor $dn/dT$ in Eq. 1.

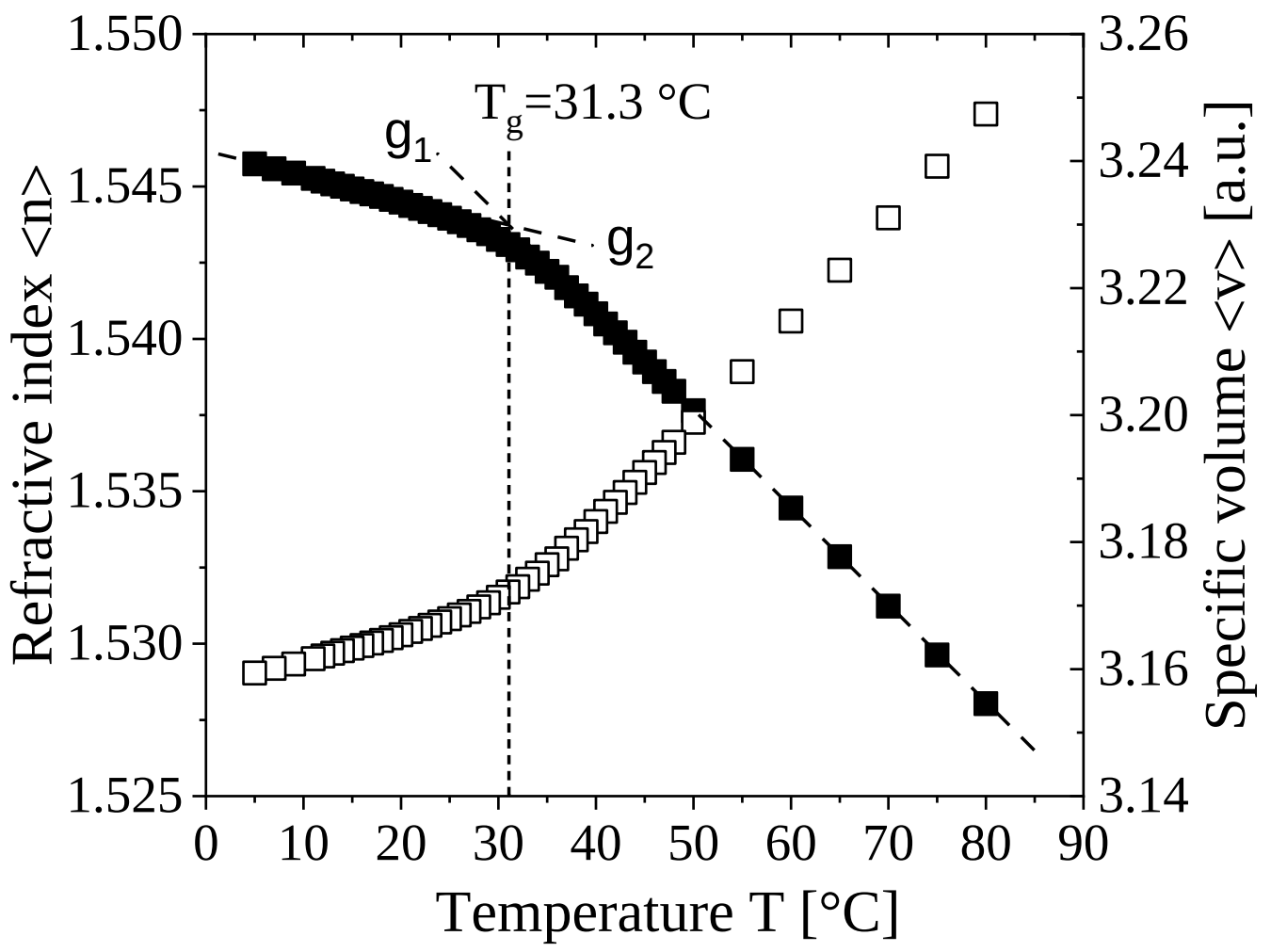

FIG 1. The short time averaged optical refractive index $\langle n \rangle = \bar{n}$ and the normalized specific volume $\langle v \rangle$ of our model epoxy (DGEBA/DETA 100/10) as a function of temperature. The so-called *operational* glass transition temperature $T_g$ was defined as usual by the intersection of high-temperature ($g_1$) and low-temperature fit lines ($g_2$) of $\bar{n}(T)$ at $T_g = 31.3$ °C.

The evaluation of the results described in Fig. 1, leads to an important initial finding: A significantly improved realization of thermodynamic equilibrium conditions in the vicinity of the canonical glass transition does not lead to a sharpening of the glass transition phenomenon on the temperature axis. Thus, the change of the refractive index $\langle n \rangle(T)$ or the specific volume $\langle v \rangle(T)$ as a function of temperature does neither become more abrupt nor kink-like.

Fig. 2 provides insights into the dynamic behaviour of thermal volume change behaviour given by $\beta^* = (\beta', \beta'')$ as simultaneously measured with TMOR, alongside the quasi-static behaviour given by $\beta$. The modulation times $\tau_{mod}$ are between $60\ s$ and $3 \cdot 10^5\ s$, i.e. the modulation frequencies $\nu_{mod}$ range from 17 $mHz$ to 3.3 $\mu Hz$.

As can be seen, in the pure visco-elastic temperature range (for $T > 55\ °C$) the thermal volume expansion coefficients $\beta' = \beta$ are independent of the modulation frequencies $\nu$. For all modulation frequencies $\nu < 17\ mHz$, the corresponding $\beta''$ values approach basic attenuation values within a margin of error. This means, the corresponding $\alpha$-relaxations have disappeared. An exception to this is the anomalous, yet physically relevant loss behaviour of $\beta''$ at 17 $mHz$, which yields a new perspective on the operational capability of the TMOR technique. This aspect will be discussed in more detail further into the text.

So far, the data described in Fig. 2 was within the expectations for dynamic investigations of the canonical glass transition. However, the data also contains a real surprise that calls into question previous interpretations of the nature of the canonical glass transition. An analysis of the loss peaks described by $\beta''(T)$ near the thermal glass transition shows that, surprisingly, their widths decrease as the modulation frequency $\nu$ decreases. This result is in clear contradiction to the attenuation behaviour predicted by the *stretched exponential* behaviour according to e.g. Havriliak-Negami[27] and Kohlrausch-Williams-Watts[28–30]. In accordance with the Kramers-Kronig relation[31,32], the accompanying $\beta'(T)$-anomaly becomes sharper as a function of temperature and decreasing modulation frequency $\nu$. In Fig. 11 in the Methods section, it is shown that such a narrowing of the relaxation time distribution takes indeed place.

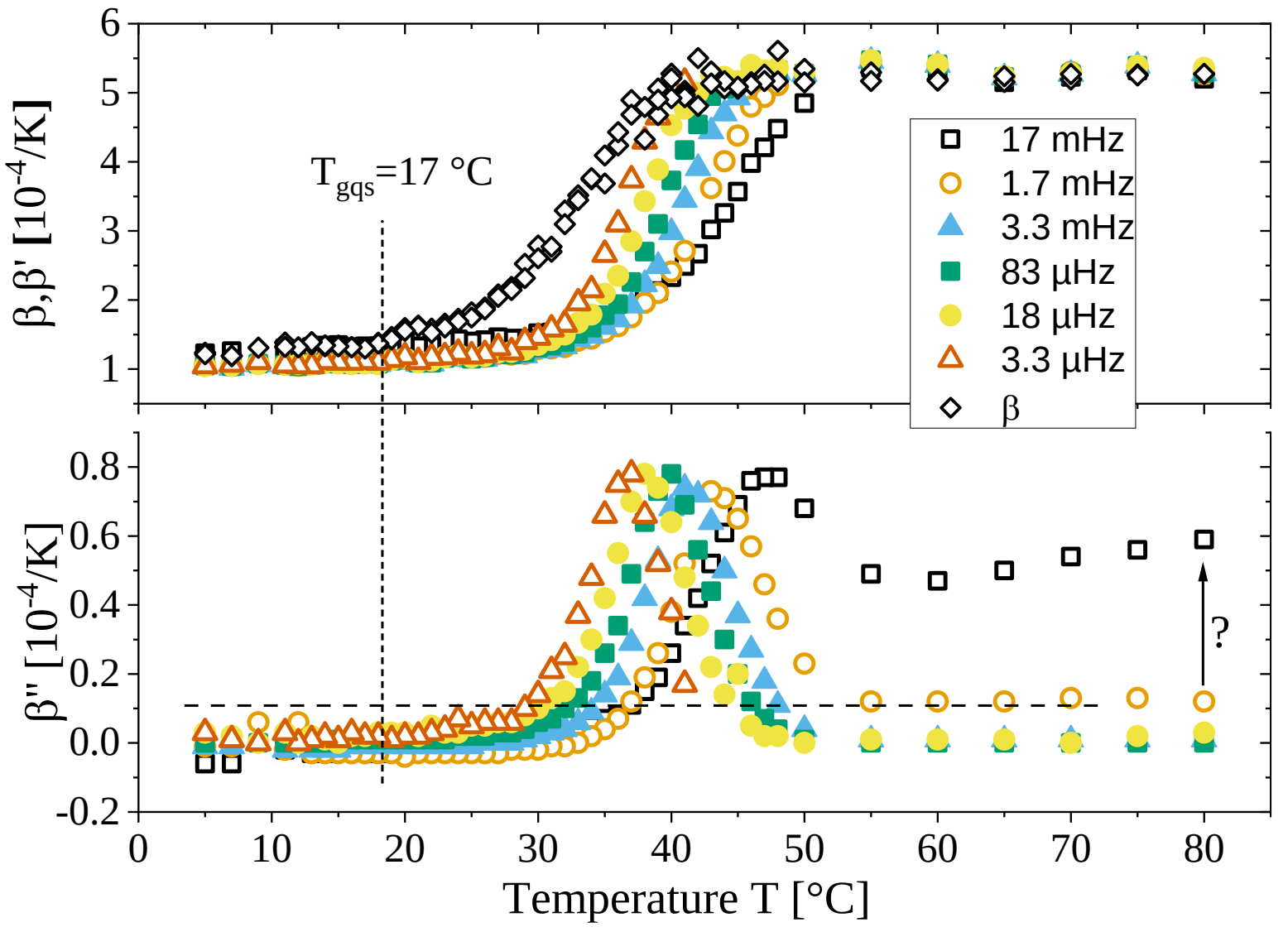

FIG 2. Static and dynamic properties of the thermal volume expansion coefficient $\beta$ and $\beta^* = \beta' + i\beta''$, respectively, of an epoxy polymer (DGEBA/DETA) as a function of the average temperature $T$ and $\tilde{T}$-modulation frequency $\nu$. The range of probed frequencies at the same temperature lies between 17 $mHz$ and 3.3 μ$Hz$ (s.a. insert). The static glass transition $T_{gqs}$~17 °$C$.

Hence, the above described behaviour of the dynamic thermal volume expansion $\beta^*(T)$ in the region of near and above the thermal glass transition raises fundamental questions. According to the theory of linearized irreversible thermodynamics[11], the relaxation time behaviour of different susceptibilities is generally not identical, i.e. strong differences between them are expected. However, the construction of VFT curves based on different dynamic measurement variables (e.g. $c^*_{ijkl}$, $\varepsilon^*_{ij}$, $c^*_p$) is quite common[33] and therefore violates the theoretical boundary conditions. The dynamic behaviour found in Fig. 2 contradicts the predictions in the polymer literature and corresponds rather to the theoretical requirements of irreversible thermodynamics. This raises the key question whether dynamic thermal volume expansion plays a role distinct from other measured variables in the vicinity of the thermal glass transition, or whether there are other reasons for the seemingly anomalous behaviour of this physical property. In fact, there could be a completely different explanation for the observed, anomalous behaviour of the thermal volume expansion coefficients $\beta(T)$ and $\beta^*(T)$ shown in Fig. 2: If the anomaly (jump) in the quasi-static thermal volume expansion $\beta$ presented in Fig. 2 has origins that are independent of collective / cooperative molecular dynamics, i.e., morphological causes, then these could influence the dynamic behaviour of thermal volume expansion by restricting/sharpening the associated relaxation behaviour. In this context it should be noted that most of the dynamic parts of the $\beta'$- and $\beta''$-curves produced by TMOR lie below the $\beta$-curve. This hypothesis involves a novel additional mechanism which narrows the line widths of the $\beta''(T)$-peaks ($\alpha$-process) in the immediate vicinity of the thermal glass transition.

So far, only the *operative* glass transition temperature has been introduced in this work. Otherwise, only the unspecific terms "canonical glass transition" and "thermal glass transition" have been used as synonyms for other glass transition temperatures without specifying the temperatures themselves. Fig. 2 can be used to remedy this shortcoming.

It is common practice to assign the temperatures $T_{g,dyn}(\nu)$ to the maximum positions of the loss curves ($2\pi \cdot \nu \cdot \tau = 1$). These dynamic glass transition temperatures obviously depend on the modulation frequency $\nu$. This definition of $T_{gdyn}$ is unambiguous, whereas the definition of a hypothetical static or quasi-static glass transition temperature is much more arbitrary. As noted above, the combined TMOR-temperature jump measurements described here only provide access to a quasi-static glass transition temperature $T_{gqs}$. According to Fig. 2, the quasi-static glass transition temperature $T_{gqs}$ should be defined by the temperature at which the dynamically measured $\beta'(T)$-curves and the statically measured $\beta(T)$ curve merge. This definition of the quasi-static glass transition temperature $T_{gqs}$ assumes that the dynamically clamped expansion coefficient $\beta'^{\infty}$ reflects approximately the static thermal expansion coefficient $\beta(T = T_{gqs})$ at the glass transition.

An additional interesting glass transition temperature can be extrapolated based on $\beta''(T, 17\ mHz)$ performed by TMOR[34]. According to Fig. 2, $\beta'(T)$ exhibits relaxed behavior at $\nu = 17\ mHz$ and temperatures above $T > 50\ °C$. In contrast, $\beta''(T)$ increases linearly and significantly in the same temperature range. From the perspective of relaxation dynamics, the Kramers-Kronig relation[31,32] appears to be violated. The cause of this behaviour is the occurrence of an additional loss process that has nothing to do with the $\alpha$-relaxation process. This additional loss process is attributed to thermally induced friction between the prism of the TMOR refractometer and the measurement sample and disappears at lower temperatures. The mechanism is described in more detail in a recent publication[34] as well as in the Methods section (cf. Fig. 12). For the present publication, it is important to note that the TMOR method, in addition to the properties of thermal volume expansion, demonstrates that shear losses at the prism interface lead to information about shear properties of the sample that are important for understanding its behaviour at the canonical glass transition.

In Fig. 3, the disappearance of shear losses at temperatures far into the glassy state at a temperature $T_{g,dyn}^{exp}$ become visible. This static glass transition has purely dynamic causes, as it is extrapolated from a high temperature range at which the $\alpha$-relaxation process is not yet superimposed by morphological changes. The temperature difference between $T_{gqs}$ and $T_{g,dyn}^{exp}$ is approximately $30\ K$ (cf. Fig. 2). This behaviour is very reminiscent of the Vogel-Fulcher-Tamman (VFT)[23–25] behaviour of the $\alpha$-relaxation process. Although the temperature $T_{gqs} = T_0$ (VFT) and the temperature $T_{g,dyn}^{exp}$ are generated from different physical data, both show the dynamic behavior during glass formation and extrapolate on temperatures that cannot be measured. It is therefore quite possible that the presented TMOR investigations reflect the VFT-relaxation time divergence. However, whether and to what extent the temperature $T_{g,dyn}^{exp}$ is related to the VFT temperature $T_0$ is an interesting but at the moment unresolved issue.

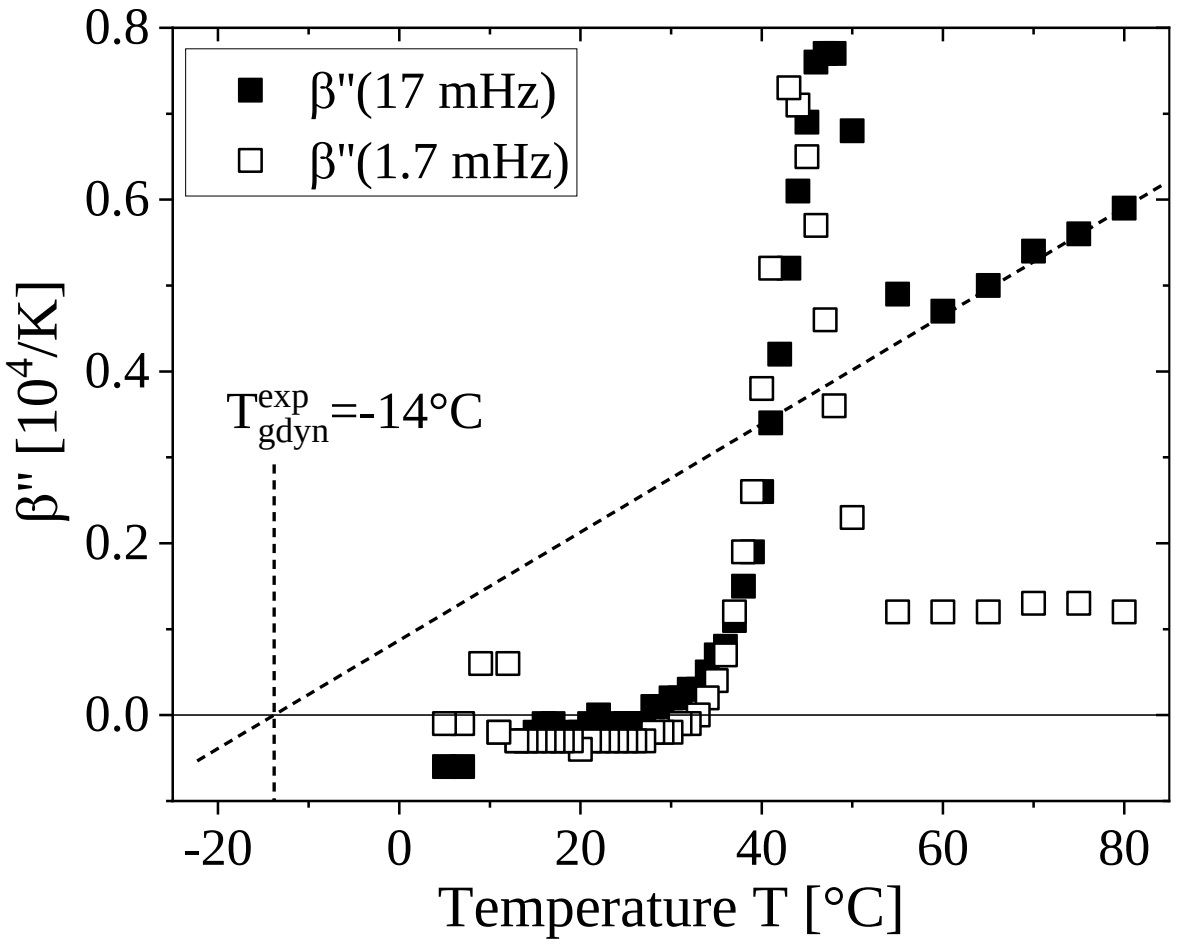

FIG. 3 Extrapolation of the quasi-static glass transition temperature $T_{g,dyn}^{exp}$ derived from shear mode excitation by TMOR at 17 $mHz$.

The TMOR data presented in this publication raises another key question: To what extent are the static and dynamic thermal volume expansion coefficients $\beta$ and $\beta^*$ are linked, and to what extent do they influence each other? Fig. 4 superimposes two dynamic measurements of $\beta'$, measured at 17 $mHz$ and 3.3 μ$Hz$, with the static $\beta$ data, simultaneously determined by the temperature jump technique. The dynamic and static data are measured simultaneously in the temperature range of the glass transition. As modulation frequencies $\nu$ for the dynamic $\beta'$-measurements the edge frequencies of the investigated frequency interval were chosen (cf. Fig. 2). The total time of TMOR investigations of the glass transition of our model epoxy was almost 2 years.

The two frequency-dependent measured thermal volume expansion coefficients in Fig. 4 clearly confirm that the jump of $\beta'$ at the dynamic glass transition becomes more spontaneous with lowering the modulation frequency, that is, on approaching static measurement conditions.

In addition to the two $\beta'(T)$-curves, Fig. 4 contains a $\beta(T)$-curve measured under static measuring conditions. In the context of TMOR and the temperature jump approach[16–18,21], the genesis of this curve requires a detailed explanation: After each temperature jump the setting of the thermal equilibrium is waited for. Then the frequency-dependent measurements are scanned one after the other. Every second a short time averaged refractive index is projected in addition to a ($\beta'$, $\beta''$)-data pair. All these refractive index data are averaged and the long-term average is assigned to the measurement temperature. The discrete interpolation points ($\langle n \rangle$,$T$) determined in this way are connected by means of a cubic spline to a continuously differentiable curve. From this $n = n(T)$ curve, the static expansion coefficient $\beta$ is determined using Eq. 1.

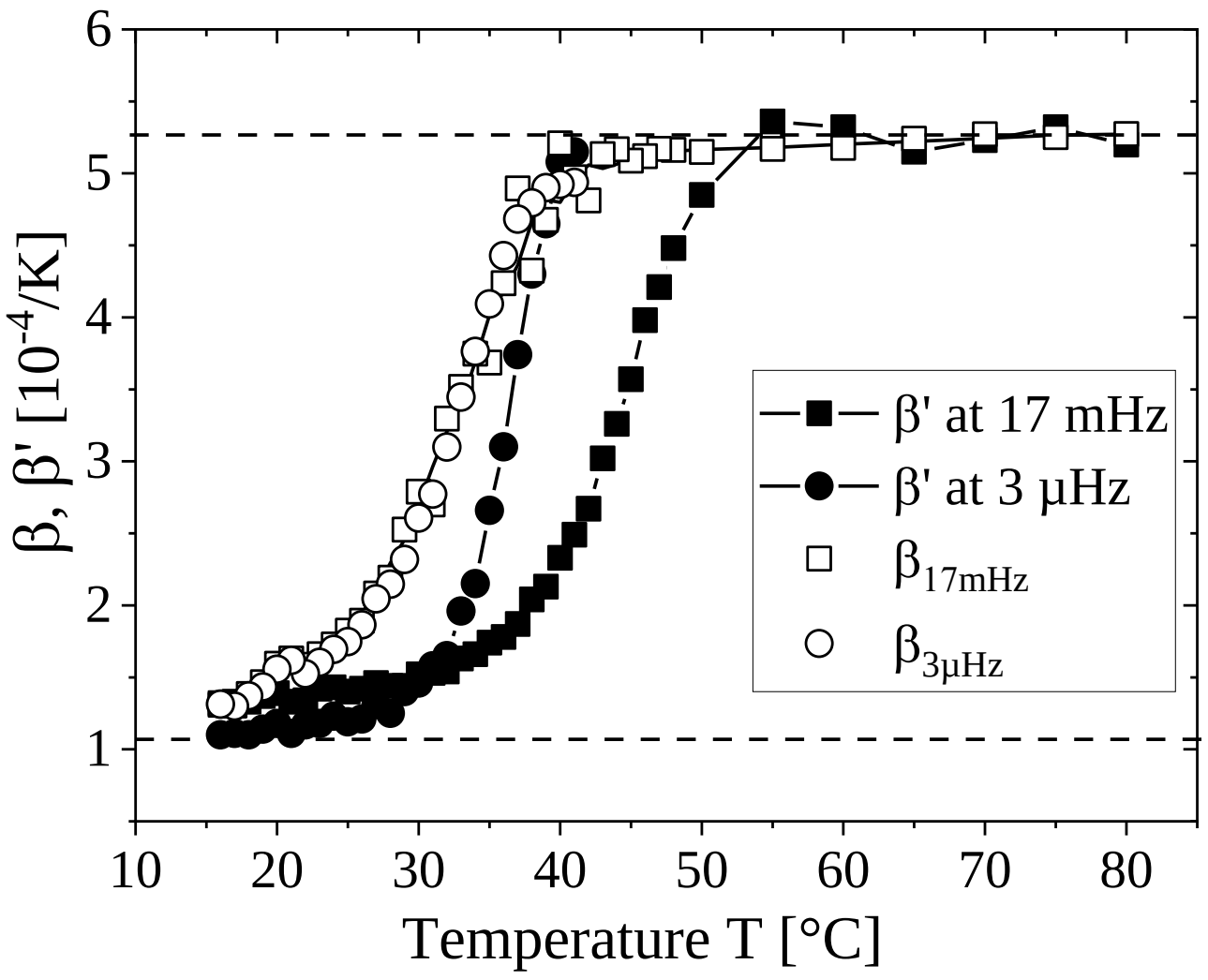

FIG 4. Static and dynamic thermal volume expansion coefficients $\beta$ (Eq. 1 and Ref.[13]) and $\beta'$ measured at 17 mHz and 3 μHz.

The fact, that the slope of the jump of $\beta'(3.3\mu Hz)$-curve is much larger than that of the static $\beta$-curve leads to the conclusion that the habitus of the dynamically measured curves for decreasing frequencies does not converge against the habitus of the statically measured curve. This provokes the question whether the physical origins leading to the static $\beta$-curve are not the same as those causing the dynamic behavior of $\beta'(\nu, T)$. The discrepancy between $T_{gqs}$ and $T_{g,dyn}^{exp}$, Fig. 2 and 3, respectively, supports the finding that the quasi-static glass transition $T_{gqs}$ has largely independent causes from the dynamically induced glass transition at $T_{g,dyn}^{exp}$. This independence of the dynamic from the static glass transition phenomenon clarifies why the experimental dynamic glass transition temperature $T_{g,dyn}^{exp}$ can only be obtained through extrapolation; the molecular dynamics of the $\alpha$-process become previously truncated by the quasi-static glass transition.

*The significant improvement of the thermodynamic equilibrium conditions in the vicinity of the canonical glass transition of our model epoxy system has provided us with reliable thermal volume expansion data for characterizing the glass transition and the quasi-static glass transition temperature $T_{gqs}$. These data indicate the existence of a static glass transition anomaly, which, however, does not occur as a sharp spontaneous event as a function of temperature. A completely unexpected result of this investigation is that the static $\beta$-anomaly is responsible for an unexpected line width reduction of the damping maxima $\beta''(\nu)$ as the glass transition is approached, while their frequency shift continues to exhibit VFT behavior. This static anomaly of the thermal volume expansion coefficient $\beta$ is also the reason for a dynamically induced glass transition at $T_{g,dyn}^{exp}$, which cannot be measured but only be extrapolated. Finally, the overall evaluation of all TMOR data in this publication suggests the hypothesis that the cause of the canonical glass transition in polymers is in fact a phase transition. According to literature the so-called "random first order theory" (RFOT) of Wolynes and coworkers[10,35,36] offers such an independence of the dynamic susceptibility $\beta'$ and the order parameter susceptibility $\beta$.*

# Methods

## Material:

The model polymer system in this study is a percolated polymer based on a mixture of Diglycidyl ether of bisphenol A (DGEBA, D.E.R. 331) and Diethylenetriamine (DETA). The mixing ratio of DGEBA and DETA was 100 to 10 parts per weight, in order to create a polymer network that has a glass transition temperature well in the accessible temperature range of TMOR, $T_g \sim 31\ °C$. The liquid mixture was cured for 1 week at room temperature and post-cured at 80 °C for another 3 weeks.

## Temperature-jump approach:

The temperature jump approach was first implemented in the field of Brillouin spectroscopy to investigate low-frequency relaxations of the phonon spectrum near the canonical glass transition in model polymers using time-domain spectroscopy[16]. A schematic representation of the procedure is shown in Fig. 5.

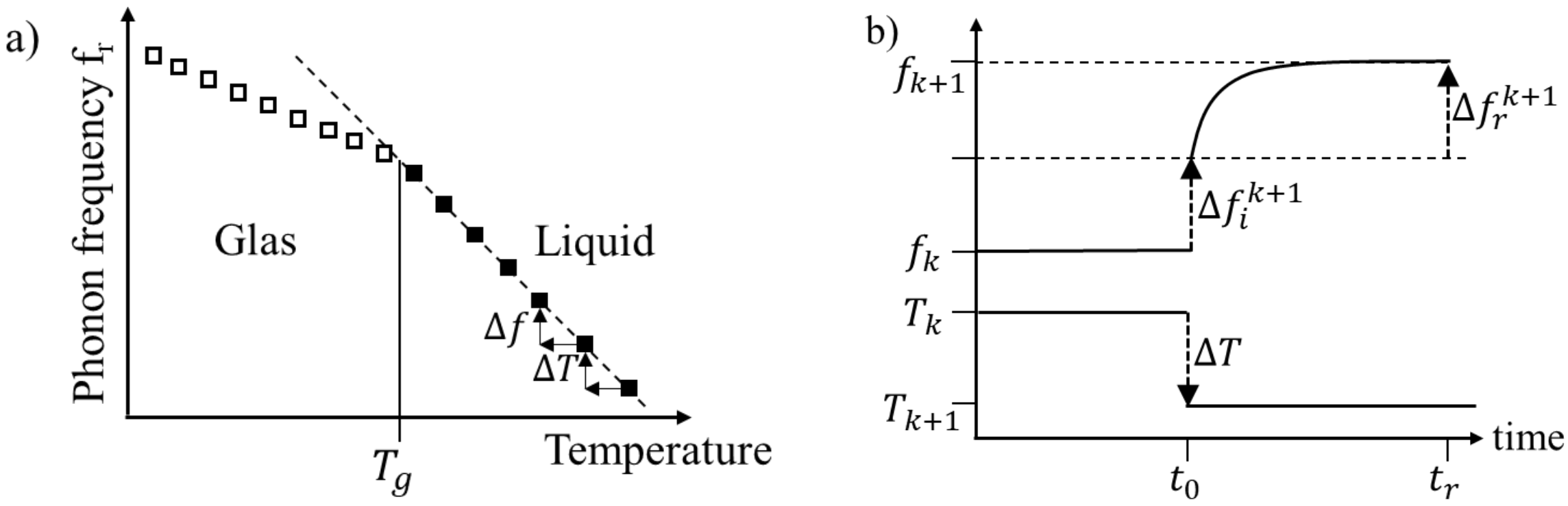

**FIG 5.** Schematic representations of (a) the equilibrated phonon frequency $f_r$ as function of temperature $T$, (b) the $k+1$ temperature jump ΔT on cooling and the resulting temporal response in the hypersonic frequency range $f$. $t$ is the time, $t_0$ indicates the time of the temperature jump, the indices $i$ and $r$ stand for the instantaneous and the relaxing part of the frequency response, respectively.

Since Brillouin spectroscopy measures thermal density fluctuations[16], no additional external disturbance is required to determine the sound frequency $f$.

Fig. 6 shows a typical temperature jump measurement for a classical model glass former, an amorphous polyvinyl acetate (PVAc). Just 10 °C above the glass transition ($T_{g,stat}{\sim}21\ °C$), the main relaxation time $\tau$ (red fit data obtained through the Kohlrausch-Williams-Watts (KWW) function[28,29], cf. Eq. 4) reached a value of 9600 s, already.

$$\Phi(t) = \frac{f_{j+1}^{\infty} - f(t)}{f_{j+1}^{\infty} - f_{j+1}^{0}} \cdot exp\left\{-\left(\frac{t}{\tau}\right)^{\beta_k}\right\} \tag{4}$$

where $f_{j+1}^{0}$ is the start of the $j+1$-th temperature jump, $f_{j+1}^{\infty}$ is its equilibrium value after the temperature jump, $\tau$ is the main relaxation time and $\beta_k$ is the stretching exponent:

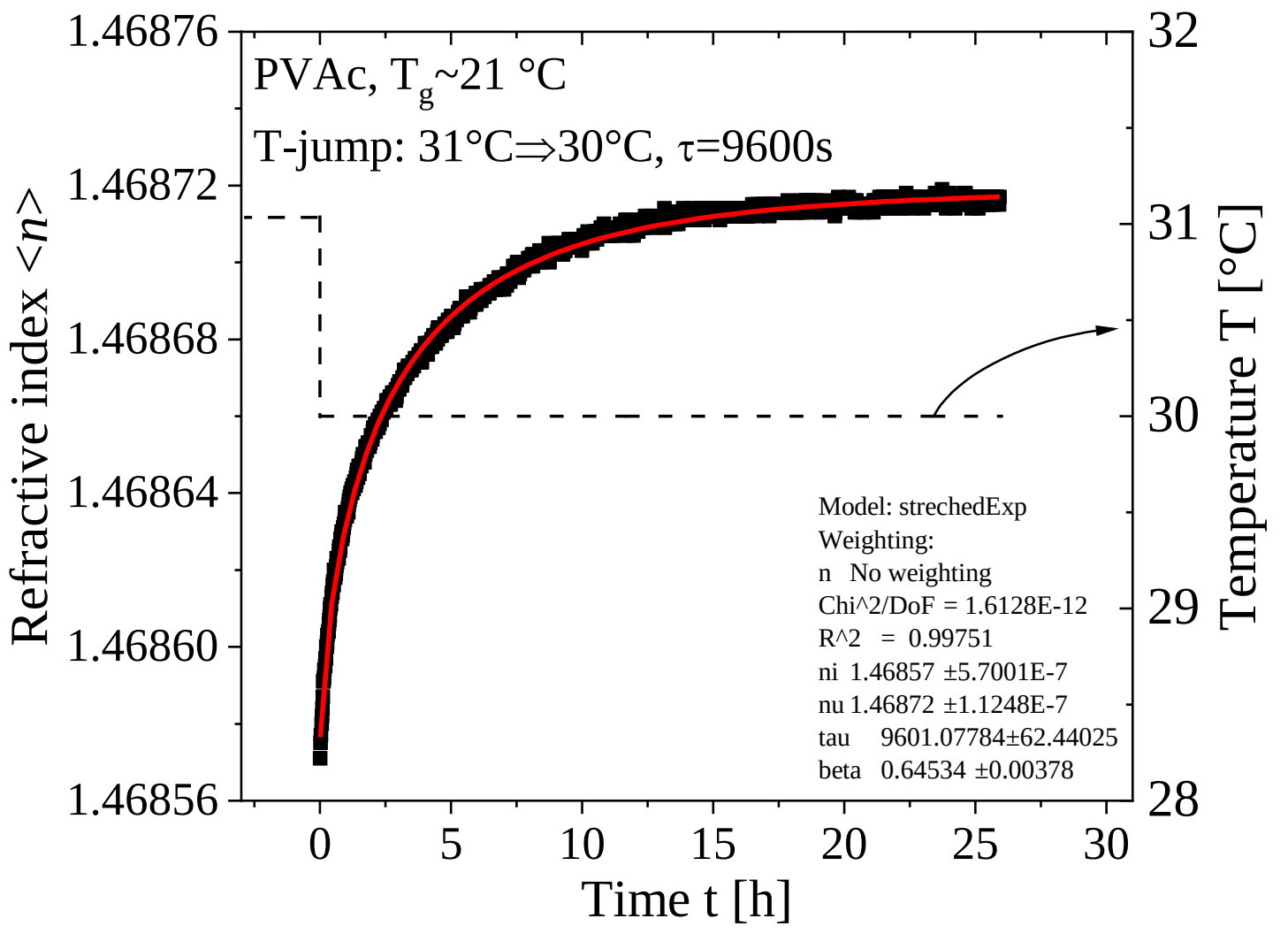

**FIG 6.** Refractive index response of PVAc after a temperature jump from 31 °C to 30 °C measured via Brillouin spectroscopy. The KWW fit is highlighted in red. The main relaxation time was $\tau = 9600s$.

Following the above described process, Fig. 7 shows the main relaxation times $\tau_\alpha$ of PVAc in the vicinity of the glass transition $T_g{\sim}21\ °C$. The data was obtained from KWW fit functions of the relaxation behavior of the refractive index response after a temperature jump to the approach-temperature. The closer the approach-temperature is to the glass transition of the material, the longer are the relaxation times. They increase in an exponential VFT-manner. 8 K above the canonical glass transition the main relaxation time $\tau$ has already reached 37.5 h. In order to avoid any undesirable nonlinear behaviour, small temperature jumps are recommended as $T_g$ is approached.

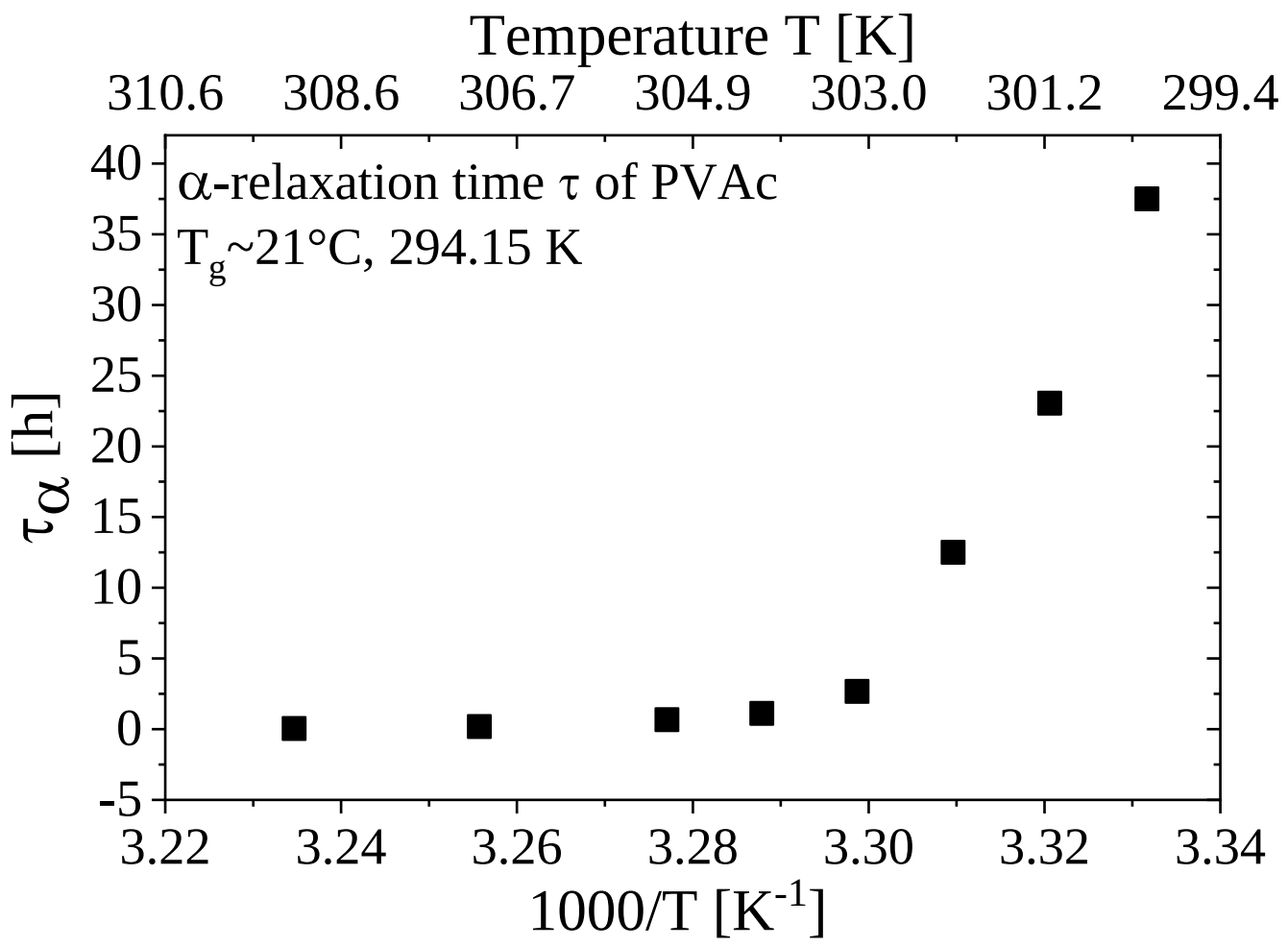

**FIG 7.** Brillouin spectroscopy measurements of relaxation times with the temperature jump method and time domain analysis.

**Temperature-jump approach with TMOR:**

In the case of TMOR experiments an external temperature disturbance, the temperature modulation $A_T$ is required to obtain a response signal of the thermal volume expansion, as is shown in Fig. 8. *Note that the time difference between the TBS and TMOR experiments are several years, and the PVAc have different molecular weights.*

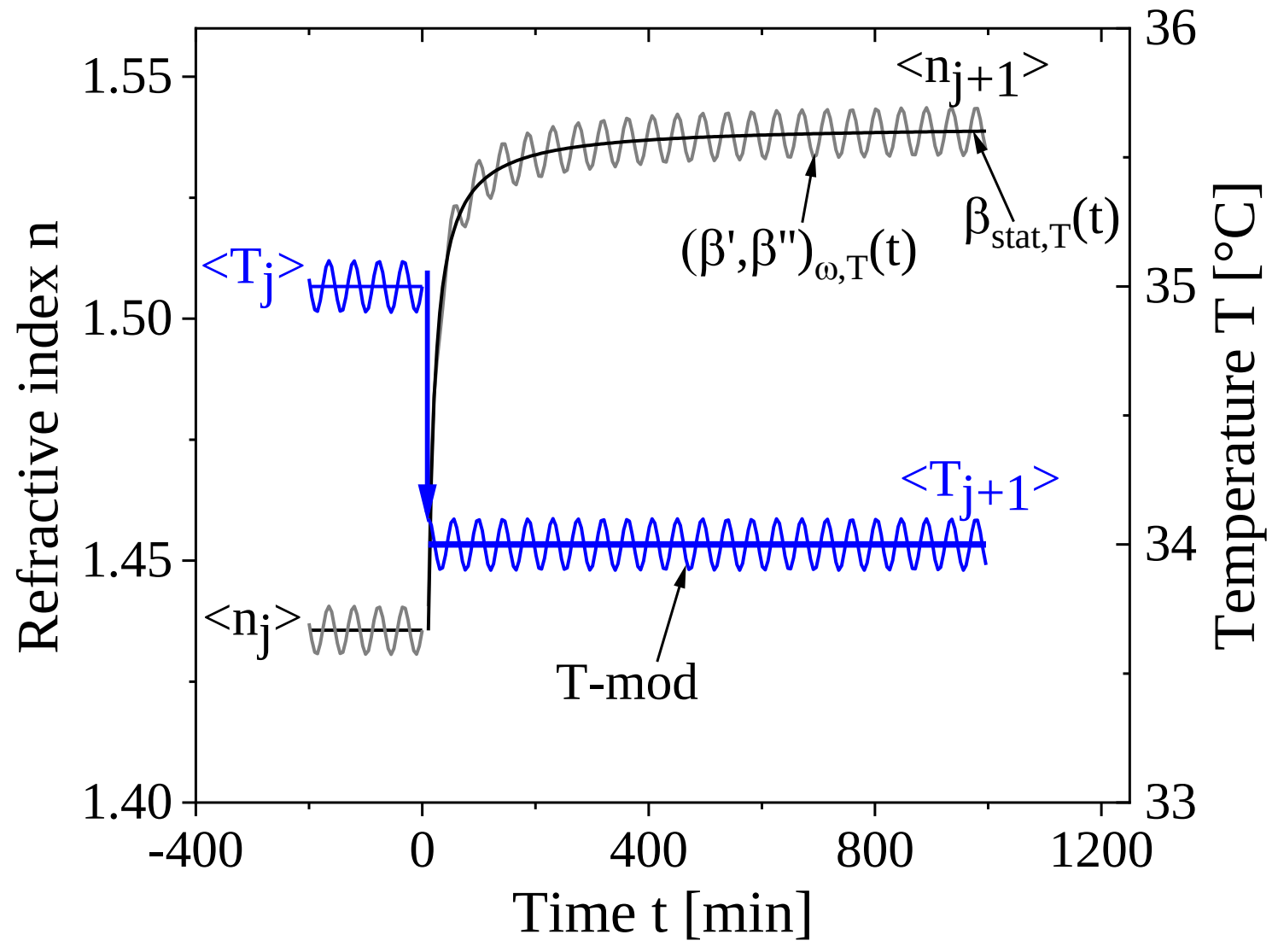

**FIG 8.** Temperature jump from $T_j$ to $T_{j+1}$ superimposed by a sinusoidal temperature modulation with $A_T = 0.1\ K$. From the average $\langle n_{j+1} \rangle$ and the complex response function $n_{j+1}^*$ one can obtain the static $\beta_{stat}$ and the dynamic ($\beta'$ and $\beta''$) thermal volume expansion coefficients, respectively, cf. *Notes on TMOR* below

Fig. 9 shows the refractive index response for PVAc ($T_{g,stat}{\sim}20\ °C$) after a temperature jump from $30\ °C$ to $29\ °C$[17]. The pronounced curvature of the $n$-response reflects the time needed for the sample to thermally equilibrate. A clear coupling between the $\alpha$-relaxation time $\tau$ and the time to thermally equilibrate the sample close to the glass transition becomes evident.

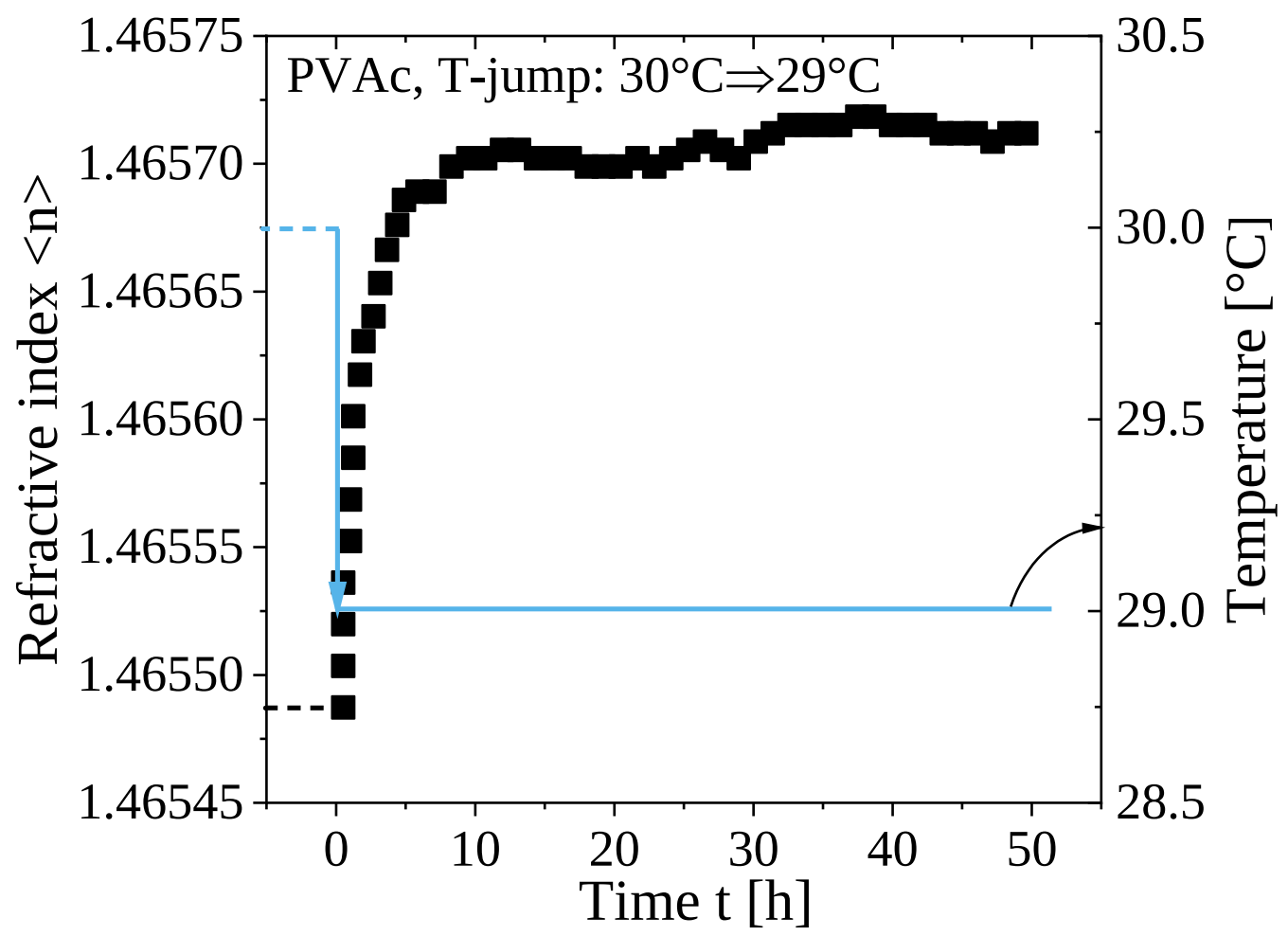

**FIG 9.** Refractive index response $n$ of PVAc after a temperature jump from 30°C to 29°C measured via TMOR, $T_{g,stat} \sim 20\ °C$

The enormous amount of time required to achieve thermodynamic equilibrium conditions near the thermal glass transition is usually ignored, leading to incorrect estimates of the physical properties at the glass transition. This is particularly true for experiments conducted at temperature rates.

**Non-linearities:**

The modulation amplitude $A_T$ of the sin-signal, schematically shown in Fig. 8, must be controlled in order to satisfy *linear response* conditions. As can be seen in Fig. 10, for the model system DGEBA/DETA, modulation amplitudes higher than approx. $A_T > 0.3\ K$ should not be applied. In fact, in the current study, the modulation amplitude was selected to be as small as $A_T = 0.1\ K$ to avoid any unintended non-linearities during the measurement.

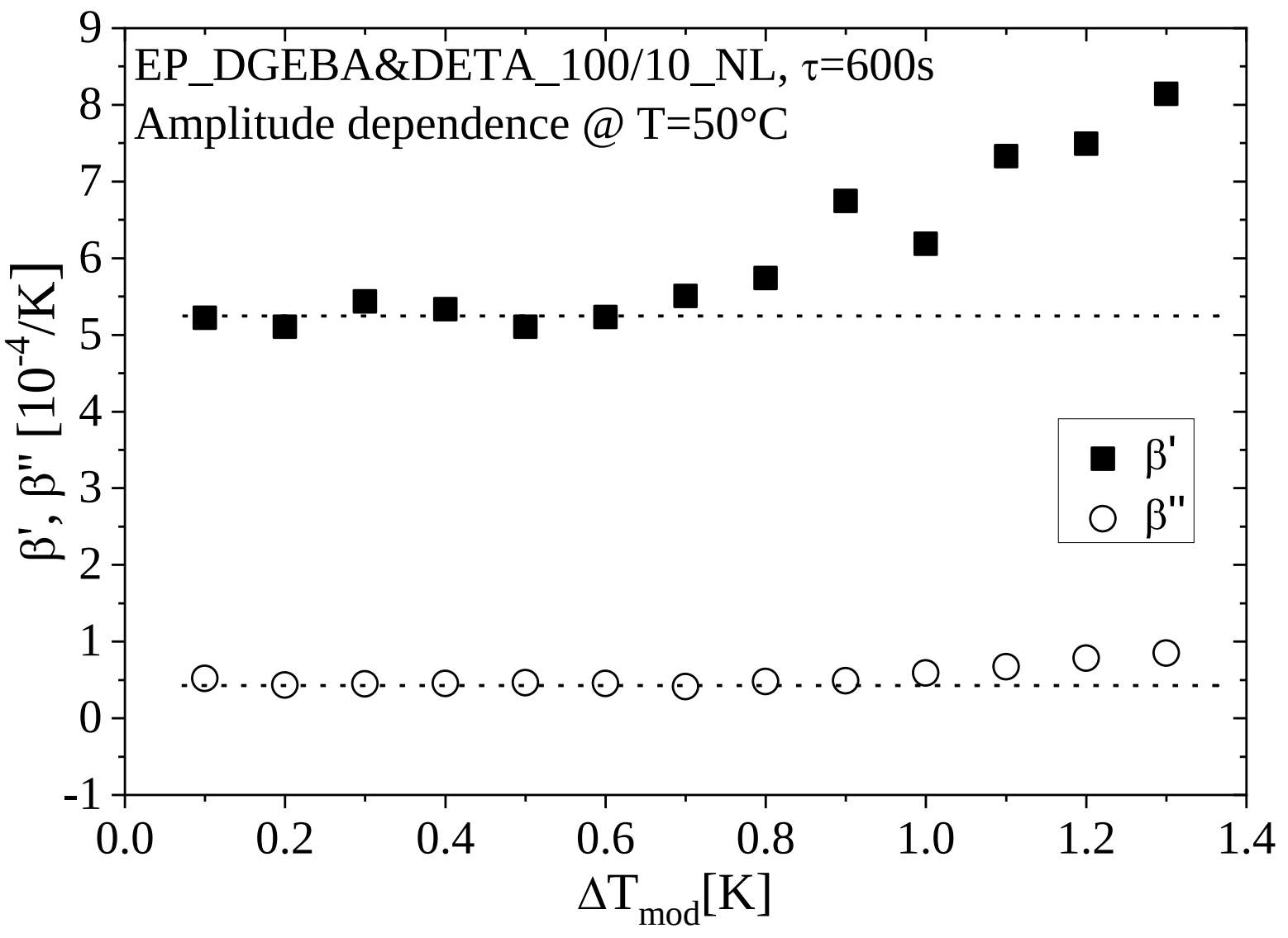

**FIG 10.** Control of the influence of temperature modulation amplitude $A_T = \Delta T_{mod}$ on the system response $(\beta', \beta'')$ of the DGEBA/DETA model system under quasi-isothermal conditions at $T = 50\ °C$.

**Some more notes on temperature-modulated optical refractometry (TMOR):**

From a theoretical point of view, the TMOR technique is based on two relationships, the Lorenz-Lorentz relation

$$\frac{n^2-1}{n^2+2} = \frac{r}{v} = r \cdot \rho, \tag{5}$$

and the thermo-optical coefficient

$$\Psi^* = \frac{dn^*}{dT} = \Psi' + i\Psi'' = \left|\frac{\partial n^*_\omega}{\partial T}\right| e^{i\Phi} = \frac{A_n}{A_T} e^{i\Phi} \tag{6}$$

where $n$ is the refractive index of the sample of interest, $r$ is the specific refractivity, $v$ is the specific volume, $\rho$ is the mass density, $\Psi$ is the thermo-optical coefficient, $T$ is the temperature, $\Phi$ is the phase angle between the sinusoidal temperature modulation and the refractive index response, $A_T$ is the amplitude of the temperature modulation and $A_n$ is the amplitude of the refractive index response to the temperature modulation.

Provided one can avoid cooling-rate influences and provided the specific refractivity $r$ can be treated as constant[2], Eq. 7 yields the static thermal expansion coefficient:

$$\beta(T) = \frac{-6 \cdot n(T)}{(n^2(T)-1)(n^2(T)+2)} \cdot \frac{\partial n(T)}{\partial T}. \tag{7}$$

The dynamic (complex) thermal volume expansion coefficient $\beta^*_{\omega=2\pi\nu}$ is obtained by combining Eq. 5 and Eq. 6 together with Eq. 8, the sinusoidal temperature perturbation:

$$T(t) = T_0 + A_T \cdot \sin(2\pi\nu t) \tag{8}$$

$$|\beta^*_\omega| = \left\{\frac{-6 \cdot \langle n \rangle}{(\langle n \rangle^2-1)(\langle n \rangle^2+2)}\right\} \cdot |\Psi^*| = \frac{-6 \cdot \langle n \rangle}{(\langle n \rangle^2-1)(\langle n \rangle^2+2)} \cdot \left\{\frac{A_n}{A_T}\right\} \tag{9}$$

with

$$\beta' \equiv Re(\beta^*_\omega) = |\beta^*_\omega| \cos\Phi \tag{10a}$$

and

$$\beta'' \equiv Im(\beta^*_\omega) = |\beta^*_\omega| \sin \Phi \quad (10b)$$

where $\langle \quad \rangle$ indicates a short time average over two frequency cycles.

It is evident that if in the course of a measurement, the losses $\beta''$ disappear, i.e., the phase angle $\Phi$ becomes zero, the static $\beta$-value is measured.

The modulation frequencies that can technically be achieved with TMOR are roughly between 1 $\mu Hz$ and 10 $mHz$. The upper frequency limit is determined by the thermal conductivity of the prism as well as the sample. The lower frequency limit is defined by the long-term stability of the TMOR measuring devices.

In this regard, it should be noted in particular that a special feature arises from Eq. 8 and 9. Apart from the temperature disturbance $A_T$, no further temperature disturbance is required to determine the thermal volume expansion coefficient. In other words, TMOR can be used to measure the thermal volume expansion coefficient under quasi-isothermal conditions at a temperature $T_0$.

**Narrowing of the relaxation time distribution on approaching the glass transition:**

In Fig. 11, a narrowing of the relaxation time distribution is shown. The superposition of the black solid curve (based on an interpolation of the data set obtained at 1.7 mHz) with the circular unfilled data points ($\nu = 3.3$ $\mu Hz$) clearly demonstrates the narrowing of the relaxation time distribution with decreasing modulation frequency $\nu$.

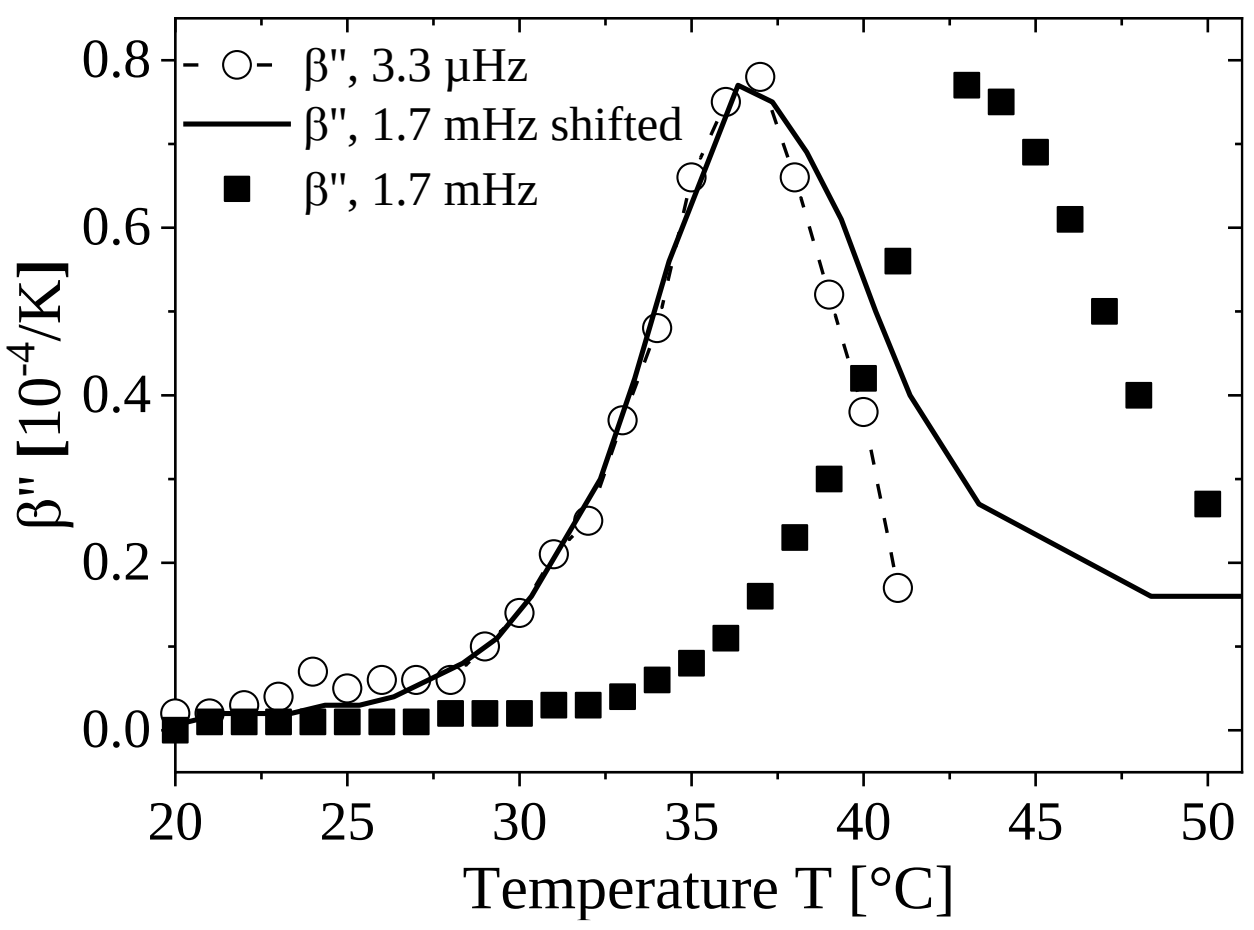

**FIG 11.** Imaginary part $\beta''(\nu)$ given as a function of temperature $T$ for the two modulation frequencies $\nu = 1.7\ mHz$ (filled squares) and $\nu = 3.3$ µ$Hz$ (unfilled circles, interpolation dashed lines). The black line represents an interpolated dataset obtained from the 1.7 $mHz$ measurement that has been shifted on the T-axis for comparison.

**High temperature shear losses as a perspective to a static glass transition:**

Fig. 12 provides a shear-loss-induced perspective on the static glass transition at the example of a different epoxy system[34]. For $\nu > 1.7\ mHz$, the sample response to the temperature modulation causes a friction event at the sample-prism surface that seemingly vanishes when the material becomes essentially frozen At $T_{g,TOC}$ and the shear modes get lost.

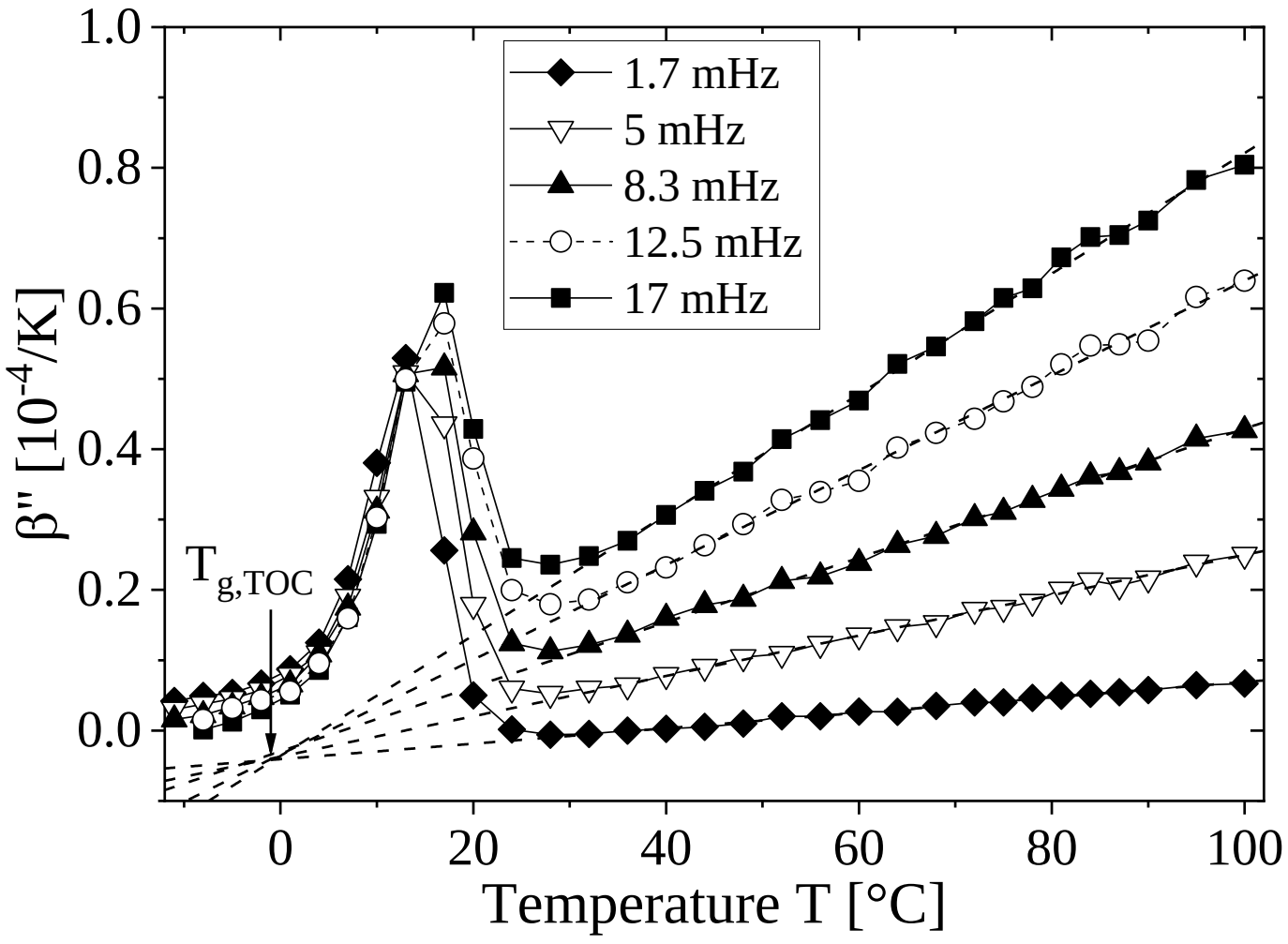

**FIG 12.** Extrapolation of shear losses of an epoxy resin that has been cured with fatty acids[34],

$$T_{g,TOC} = T_{g,dyn}^{exp} \sim -1°C$$

## DATA AVAILABILITY STATEMENT

Upon publication, the data will be made publicly available on the European open data repository Zenodo.org

## ACKNOWLEDGEMENTS

The authors gratefully acknowledge the discussions with Martine Philipp and Wulff Possart to this work. Also, the authors want to emphasize the outstanding collaboration with Anton Paar OptoTec GmbH, especially with Martin Ostermeyer.

## AUTHOR CONTRIBUTIONS

**JKK** performed the investigations, conceptualized and wrote the initial draft of the manuscript, **AK** curated and visualized the data, **AK and BW** edited and revised the initial draft.

## COMPETING INTEREST DEECLARATION

The authors declare no competing interests.

## ADDITIONAL INFORMATION

The are no supplementary data attached to this manuscript.

## CORRESPONDING AUTHORS

Jan-Kristian Krüger: jan-kristian.krueger@ivw.uni-kl.de

Andreas Klingler: andreas.klingler@ivw.uni-kl.de